\documentstyle[12pt]{article}
\newcommand{\Kp}{$K^+$}
\newcommand{\Ko}{$K^0$}
\newcommand{\Km}{$K^-$}
\newcommand{\Kbo}{$\bar K^0$}
\newcommand{\SigKN}{ \mbox{$\Sigma_{KN}$} }
\newcommand{\sigm}{\mbox{$\langle \sigma \rangle$}}
\newcommand{\psiN}{\mbox{$\psi_N$}}
\newcommand{\psiL}{\mbox{$\psi_\Lambda$}}
\newcommand{\gs}{\mbox{$g_\sigma$}}
\newcommand{\gv}{\mbox{$g_\omega$}}
\newcommand{\mv}{\mbox{$m_\omega$}}
\newcommand{\gsL}{\mbox{$g_\sigma^\Lambda$}}
\newcommand{\gvL}{\mbox{$g_\omega^\Lambda$}}
\newcommand{\vp}{\mbox{\boldmath $p$}}

\newcommand{\eq}{\label}

\newcommand{\stret}[1]{\hbox{$\vcenter to #1{}$}}

\begin{document}

\hfill{KUNS-1400}

\begin{center}

{\Large \bf  
\Kp(\Ko)- Condensation in Highly Dense Matter 
with the Relativistic Mean-Field Theory }

\vspace{1cm}

Tomoyuki Maruyama, \\[0.5ex]
$^1$Advanced Science Research Center,\\
Japan Atomic Energy Research Institute,\\ 
Tokai, Ibaraki 319-11\\[2ex]
Hiromasa Shin, Hirotsugu Fujii, \\
and  Toshitaka Tatsumi \\[0.5ex]
$^2$Department of Physics, Kyoto University,\\ Kyoto 606-01, JAPAN

\vspace{1cm}

\begin{abstract}
Properties of dense hadronic matter including strange particles are
studied within the relativistic mean-field theory (RMFT). The possibility of
kaon condensation is reexamined, and a simple condition is found for 
the parameters included in RMFT.   

\bigskip\noindent
PACS numbers: 13,75.Jz, 13,75.Ev, 21.65.+f, 21.80.+a, 24.10.Jv, 25.75.+r

\noindent
Key Words: strange hadronic matter, reletivistic mean-field theory,
kaon condensation
\end{abstract}

\bigskip

\end{center}

\newpage

In relativistic heavy-ion collisions,
it is expected that the hot and dense zone 
includes many species of hadrons, where  
a lot of strange particles should be produced.
Recent studies have shown that kaon dispersion relation would be much 
changed in nuclear medium, which gives rise to kaon condensation in 
neutron star matter \cite{rev} on one hand and 
modification of kaon production in heavy-ion collisions \cite{kol}, 
change of the dilepton production 
rate from RBUU \cite{LiKo} or the fire ball \cite{Tatsu} on the other hand.
For the former subject, the hyperon degrees of freedom has been
extensively studied 
by way of the relativistic
mean-field theory (RMFT) \cite{ell,sch}. 
In this letter we consider strange hadronic matter by extending RMFT to 
incorporate nucleon, $\Lambda$ hyperon and
$K$ meson,  which are essential degrees of freedom in strange
hadronic matter. First preliminary report on this subject was given in
ref.\cite{shin}. Here we examine the relevance of the parameters
included in RMFT in more detail
and present a relation between them to give a condition for kaon
condensation in high-density matter. 

Some years ago a possibility of strange hadronic matter produced in
relativistic heavy-ion collisions was indicated
in RMFT, where abundance 
of lambda or other hyperons overwhelms that of nucleons \cite{Rufa}. 
They, however, considered only baryons. On the other hand kaons, 
the lightest strange
mesons, may also carry strangeness and they are much
modified in the medium. So it needs to take kaons into account properly  
as well as hyperons to explore strange hadronic matter.

Nelson and Kaplan \cite{Kpcon} have suggested \Kp\Km(\Ko\Kbo)-pair 
condensation in relativistic heavy-ion collisions 
\footnote{The possibility of kaon condensation in relativistic
heavy-ion collisions has been also 
suggested for a different reason \cite{cle}. }.
As increasing density the lowest energy of \Kp(\Ko) is reduced by
the $KN$ s-wave interaction (mainly by the KN-sigma term \SigKN), 
and eventually becomes equal to the strangeness chemical
potential $\mu_s(=\mu_K)$, where the \Kp(\Ko) effective mass reachs 
$-\mu_K$ simultaneously and \Kp\Km(\Ko\Kbo)-pair condensation occurs.
Their idea is very intersting, but they considered only nucleon matter 
and did not take into account any hyperon degrees of
freedom. Moreover, their result seems unlikely in light of the subsequent
studies about kaon in medium; the \Kp  excitation energy receives a repulsive
effect instead of an attractive one \cite{rev}.

In relativistic heavy-ion collisions hyperons and kaons
are produced in pairs; this
possibility is more favorable than that of the kaon-antikaon pair 
production due to
the different threshold energy.
In previous papers \cite{maru-K,fujii}, furthermore, it has been
shown that 
the relativistic effects moderate kaon condensation by the 
$self$-$suppression$ mechanism in neutron-star matter.
This self-suppression effects should also appear in symmetric nuclear matter;
the main effect caused by the $KN$-sigma term \SigKN is weakened by the small
nucleon effective mass.   

Here we study strange hadronic matter at high-density but rather low
temperature ($T<m_\pi$) by dealing
hyperon and kaon equally within RMFT. 
In particular we pay attention to the abundance of $K^+(K^0)$ which is
measured by the strangeness chemical potential $\mu_s$ (=$\mu_K$), and 
discuss the possibility of \Kp(\Ko) condensation. 
We, hereafter, restrict ourselves to hadron matter in thermal equilibrium
, and treat it at {\it zero} temperature for simplicity
because the nuclear properties such as the nucleon and meson effective masses 
are not largely changed below $T \sim m_\pi$ \cite{ft}, and thermal-pion 
effects hardly modify our conclusion. We shall give some comments on
this matter at the end of this letter, where we also briefy discuss the
observability of kaon condensation in the realistic situation of 
heavy-ion collisions.

First we  briefly explain our basic formulation.
The lagrangian density with nucleon, lambda and kaon is given as follows:
\begin{eqnarray}
{\cal L} & = & 
\bar{\psiN} (i{\not \partial} - M_N )\psiN + \gs \bar{\psiN} \psiN \sigma
+ \gv \bar{\psiN} \gamma_\mu \psiN \omega^\mu
\nonumber \\
&&+ \bar{\psiL} (i{\not \partial} - M_{\Lambda} )\psiL 
+ \gsL \bar{\psiL} \psiL  \sigma
+ \gvL \bar{\psiL} \gamma_\mu \psiL \omega^\mu
\nonumber \\
&&- {\widetilde U} [\sigma] 
+ \frac{1}{2} m_{\omega}^2 \omega_\mu \omega^\mu
\nonumber \\
&& + \{ ( {\partial_\mu} - \frac{3i}{8f^2} \bar{\psiN} \gamma_\mu  \psiN ) 
{K^\dagger} \}  
\{ ( {\partial^\mu} + \frac{3i}{8f^2} \bar{\psiN} \gamma^\mu  \psiN ) K \} 
- {m_K^2}{K^\dagger}K
\nonumber \\
&&+ \frac{\SigKN}{f^2} \bar{\psiN} \psiN K^{\dagger} K , 
\end{eqnarray}
where $\psiN$, $\psiL$ and $K$ are the nucleon, lambda and kaon fields, 
respectively, and ${\widetilde U} [\sigma]$ is the self-energy potential 
of the scalar mean-field as
\begin{equation}
\widetilde{U} [\sigma] 
= \frac{ \stret{20pt} \frac{1}{2} m_s^2 \sigma^2 
+ \frac{1}{3} B_\sigma \sigma^3
+ \frac{1}{4} C_\sigma \sigma^4 } 
{ \stret{20pt} 1 + \frac{1}{2} A_\sigma \sigma^2 } \  .
\eq{sigself}
\end{equation}

In the above expression the interaction between nucleon and kaon is given by 
the $KN$~sigma term \SigKN and the Tomozawa-Weinberg type vector interaction.
The latter interaction is introduced by the minimal coupling as in 
the usual chiral perturbation theory.
Furthermore we omit the interaction between kaon and $\Lambda$
because we treat only the system 
where the number of $\Lambda$ hyperons is very small.
In addition, the isospin-dependent terms are omitted because only the isospin
saturation system is treated here.

>From the Euler-Lagrange equations the nucleon and lambda effective masses are 
given as
\begin{eqnarray}
M_N^* & = & M_N - U_{s}(N)
\nonumber \\
M_\Lambda^{*} & = & M_\Lambda - U_{s}({\Lambda})
\nonumber \\ \nonumber \\
 m_K^{*} & = & \sqrt{ m_K^2 - \frac{\SigKN}{f^2} \rho_s(N) },
\label{efmas}
\end{eqnarray}
with the scalar potentials,
\begin{eqnarray}
U_{s}(N)& = &  \gs \sigm +  \frac{\SigKN}{f^2 m_K^*} \rho_s(K) ,
\nonumber \\
U_{s}(\Lambda) & = &  g_s^{\Lambda} \sigm, 
\label{us}
\end{eqnarray}
where the kaon scalar density $\rho_s (K)$ is equal to the
kaon number density $\rho_K$ at zero temperature, and
other scalar densities $\rho_s (\alpha)$ $(\alpha=N, \Lambda)$ 
are defined by
\begin{equation}
\rho_s (\alpha) = \frac{\gamma}{(2 \pi)^3} \int {\rm d}^3 {\vp} 
f_\alpha ({\vp}) 
\frac{M_{\alpha}^*}{ \sqrt{ {\vp}^2 + M_{\alpha}^{*2} } }   ,
\label{rhos}
\end{equation}
with the spin-isospin degeneracy factor, $\gamma = 4$ for nucleon and 
$\gamma = 2 $ for lambda, and 
the Fermi momentum-distribution $f_\alpha({\vp})$.
The scalar mean-field $\sigm$ is given by
\begin{equation}
\frac{\partial}{\partial \sigm} {\tilde U} [\sigm]
= \gs \rho_s(N)  + \gsL \rho_s(\Lambda) .
\end{equation}
Then the baryon single-particle energy is written as
\begin{equation}
\varepsilon_{\alpha}({\vp}) =  
\sqrt{ {\vp}^2 + M_\alpha^{* 2} } + U_0(\alpha),
\label{sgen}
\end{equation}
where time components of the vector potentials $U_0$ are given as
\begin{eqnarray}
U_{0}(N) &=& \frac{\gv^2}{m_{\omega}^2} \rho_N 
+ \frac{\gv \gvL}{m_{\omega}^2} \rho_{\Lambda}
+ \frac{9}{32 f^4 m_K^*} \rho_{s}(K) \rho_N
\nonumber \\
U_{0}(\Lambda) &=& \frac{\gv \gvL}{m_{\omega}^2} \rho_{N}
+ \frac{\gvL^2}{m_{\omega}^2} \rho_{\Lambda}.
\end{eqnarray}
with the nucleon density $\rho_N$ and the lambda density $\rho_\Lambda$.
 Finally, the excitation energy of kaons $\omega_K({\vp})$ can be written as 
\begin{equation}
\omega_K({\vp})=\sqrt{ {\vp}^2 + m_K^{* 2} } + U_0(K)
\end{equation}
with the Tomozawa-Weinberg vector potential, 
$U_{0}(K) = 3{\rho_N}/{8 f^2}$ for \Kp(\Ko).
It is to be noted that this
form is different from the one given in Refs.
\cite{rev,LiKo,Tatsu,maru-K,fujii}. The difference stems from how to
treat the vector interaction; this coupling scheme respects the gauge
invariance of vector interactions \cite{sch}.

If the nucleon Fermi energy $\varepsilon_{F}(N)$ becomes larger than
$M^{*}_\Lambda + U_{0}(\Lambda) + m^{*}_K + U_{0}(K)$, the $\Lambda$ and kaon
appear in matter, namely $K^+(K^0)$ condensation occurs.
Since the situation strongly depends on the value of the $\Lambda$
coupling constants $\gsL$ and $\gvL$, 
we first need to examine it in detail.

Naive $SU(3)$ symmetry relation suggests that the values of $\Lambda$
coupling constants  are two-third of nucleon ones: 
$\gsL = 2\gs/3$ and $\gvL = 2\gv/3$.
However, the coupling constants in RMFT should be different from the bare ones
because they represent the effective strength of the mean-fields including
many-body effects and higher-order correlations.
The force between two baryons is strong but short-range, so that
the mean-field cannot be perturbatively given, but its non-locality
is rarely seen in low-energy phenomena:
the $\sigma$ and $\omega$ fields are brought about by not only 
the Hartree contribution but also Fock and higher-order contributions.
For the short-range interaction such as nuclear force the non-locality 
of the Fock part is very small, and
the Hartree and Fock contributions cannot be distinguished in the 
spin-isospin saturated system at the low-energy limit.
Furthermore we have not had enough information about the interaction between
nucleon and lambda.
At present, then, we cannot determine the effective coupling constant 
$\gsL$ and $\gvL$ sufficiently.

Studies of hypernuclei have indicated that depth of the $\Lambda$ 
potential is about $28 - 30$ MeV, and that its $LS$-splitting is very small.
Boguta and Bohrmann \cite{boguta} have given $\gsL/\gs = \gvL/\gv = 0.33$ 
by fitting the $\Lambda$ single-particle level under the restriction,
$\gsL/\gs = \gvL/\gv$.
Rufa et al. \cite{Rufa}, however, have shown that the parameters 
allow more ambiguities for the case without the restriction.
\footnote{For other discussions about hypernuclei and RMFT, see
Ref.~\cite{mar} and references cited therein.} 

On the other hand Glendenning \cite{glen} have pointed out
that such small coupling constants 
too much soften EOS of neutron matter and cannot make sufficiently heavy 
neutron star with 1.5 times the solar mass.
This point must be very important to study strange hadronic matter, 
but does not
constrain the parameters so severely.
As pointed out before, the $\sigma$ and $\omega$ fields do 
include not only
the Hartree contribution but also the Fock and higher-order contributions.
In the hyperon-rich system, the Fock field between hyperons should work.
Within RMFT we should be able to stiffen EOS of strange hadronic matter
by introducing the repulsive mean-field affecting only hyperons 
such as the $\phi$- field in place of the Hartree-Fock calculation.
In this letter we do not introduce such fields because we treat only
the hyperon-poor system. 

Here we would like to suggest one more important hint for the
$\Lambda$ 
mean-field
from the optical potential analysis in proton-nucleus elastic 
scatterings \cite{KLW1,Hama}.
In order to reproduce the proton-nucleus optical potential observed 
experimentally, the strength of the vector mean-field must be
inversely proportional to the incident energy \cite{KLW1,Hama} around
1 GeV of proton incident energy.
In general only the Hartree (local) parts contribute to the nucleon field 
at the high-energy limit because the Pauli blocking does not influence
nucleon with momentum far from Fermi sea.
>From the optical potential analysis, thus, we have noticed that the Hartree
contribution to the vector mean-field is very small \cite{KLW1}.
Since lambda is not affected by the Pauli effects, we can suppose that 
the mean-field for $\Lambda$ does not receive so strong many-body effects, 
and consequently the Hartree contribution should be dominant.
Hence the $\Lambda$ potential $U_{0}({\Lambda})$ is estimated to be 
very small. 

According to the above considerations we prepare two sets of parameters for 
the coupling of $\Lambda$ to the nucleon mean-fields.
One set (L1) is that $\gsL = 2\gs/3 , \gvL = 2\gv/3$ with the $SU(3)$ symmetry
relation, and the other set (L2) is that $\gvL = 0$ and \gsL~ obtained by the
following condition.
\begin{equation}
U_s(\Lambda) - U_0(\Lambda) = \frac{2}{3} ( U_s(N) - U_0(N) ).
\label{ucent}
\end{equation}
In addition we use two kinds of parameter-sets for nucleon,
PM1 ($K = 200 MeV$, ${M^*_N}/M_N = 0.7$ at $\rho = \rho_0$)
and PM4 ($K = 200 MeV$, ${M^*_N}/M_N = 0.55$ at $\rho = \rho_0$), 
and  \SigKN = 300 MeV for kaons.
The depth of the $\Lambda$ potential given by Eq. (\ref{ucent}) is a little
larger than the experimental value 30 MeV.
However we have not known the value in the infinite-matter limit
without the surface contribution, and
the small difference does not affect a rather qualitative study 
in this letter.  

Now we calculate the density-dependence of
\begin{equation}
\Delta \mu = \varepsilon_{F}(N) - M^{*}_{\Lambda} - U_{0}(\Lambda)
- m^*_{K} - U_{0}(K) 
\end{equation}
in the no-strangeness system: $\rho_\Lambda = \rho_K = 0$.
At the critical density where $\Delta \mu =0$, a pair of $\Lambda$ and 
\Kp(\Ko) appears in
the ground state, namely
\Kp(\Ko)-condensation occurs. 
Above the critical density  
chemical equilibrium among nucleon, lambda and kaon ever holds,
\begin{equation}
\varepsilon_{F}(N) = \varepsilon_{F}(\Lambda) + \mu_K
\end{equation}
with the nucleon and lambda Fermi energies, $\varepsilon_{F}(\alpha)$, 
$\alpha=N,\Lambda$, and 
the kaon chemical potential $\mu_K = m^{*}_K + U_{0}(K)$,
 which is identical to the strangeness chemical potential $\mu_s$.
Fig. 1 shows the results with PM1-L1, PM1-L2, PM4-L1 and PM4-L2.
We add that of free lambda and kaon with PM1 (PM1-free).
In all cases kaon condensation occurs, while the critical density is
different.
In the parameter-sets with L1 the critical density is very high:
$\rho_c = 16 \rho_0$ for PM1-L1 and $\rho_c = 7.6 \rho_0$ for PM4-L1.
The former critical density is too high to be attained in relativistic
heavy-ion collisions.
In the parameter-sets with L2, however, the critical density is not very high:
$\rho_c = 6.2 \rho_0$ for PM1-L1 and $\rho_c = 3.9 \rho_0$ for PM4-L1.
We should furthermore note the difference between results 
of PM1-L2 and PM1-free;
the critical density in PM1-L2 is larger than that in PM1-free though
lambda and kaon feel an attractive field in PM1-L2.
This difference is generated only by the Tomozawa-Weinberg term.
In fact the KN-sigma term cannot much contribute to the final results
because of the $self$-$suppression$ mechanism \cite{maru-K,fujii}.

In Fig. 2 we show EOS of the normal phase and kaon condensed phase (a),
the effective masses of nucleon and kaon as the ratio to their 
bare masses (b), 
and the $\Lambda$ fraction ($\rho_\Lambda / \rho_B, 
\rho_B=\rho_N+\rho_\Lambda$), (c) for PM1-L2 and PM4-L2.
We can see that kaon condensation  reduces the total energy 
per nucleon above the critical density.
The effective mass decreases for nucleon, while increases for kaon;
the latter feature stems from the reduction of the nucleon fraction.
Although the reduction of EOS and the change of the effective masses are 
not so large, the $\Lambda$ density increases rapidly, especially for PM4-L2.
Since our model does not involve any field which works between
hyperons and the kaon-lambda interaction,
properties of the system above the threshold
are not quantitatively realistic except near the critical density. 
So we should not take these results so seriously.

Since the scalar density $\rho_s$ approaches to a finite value at the infinite
density limit in RMFT, the vector mean-field dominantly affects the 
nuclear EOS in high-density regime.
Hence the difference between the results of PM1 and PM4 comes from the 
different 
strength of the vector mean-fields.
Namely kaon condensation is brought about by the balance among
the different vector mean-fields of nucleon, lambda and kaon.
Therefore the condition for \Kp(\Ko)-condensation is given as
\begin{equation}
\frac{\gv^2}{\mv^2} - \frac{\gvL \gv}{\mv^2} - \frac{3}{8f^2} \le 0.
\end{equation}

As for the lambda potential the present parameterizations L1 and L2
should be the
extreme ones; the realistic value must be laid between two cases.
In order to clarify the ambiguity coming from the parameterization, in Fig. 3, 
we plot the relation between the critical density $\rho_c$ and 
the ratio $x_{\omega} = \gvL/\gv$ under the condition (\ref{ucent}).
The critical density dose not strongly depends on the coupling \gvL~ 
as far as  $\gvL \le 0.5 \gv$, while it steeply varies around $\gvL = 0.6 \gv$.
In our view the value of \gvL must be much smaller than L1 because of 
the small LS-splitting in lambda hypernuclei and very small Hartree 
contribution to the nucleon vector mean-field.
Hence the realistic critical density should not be so far from that in L2.

In this letter we have not discussed the effects of temperature, while 
the compressed system in relativistic heavy-ion collisions must be hot.
\footnote{We consider here the case $T<m_\pi$, since we are interested 
in moderately high-energy heavy-ion collisions around several tens
GeV/u,e.g.,the AGS
energy region \cite{AGS1,AGS2}.}
First we should note that the vector mean-field is only proportional to density
in RMFT and independent of temperature.
Furthermore it has been shown in Ref. \cite{ft} that the temperature 
dependence of the scalar mean-field is also small  below $T \sim 200$MeV.
Secondly, in a realistic situation, we should take into accout another 
temperature effect as well, where many pions may be excited. 
One may wonder whether such thermal pions significantly modify our conclusion 
through the interaction with baryons or kaons. 
However, this is not the case at temperature below the pion mass 
$(T< m_\pi)$. 
Actually the number density of pions ($\propto T^3$ in the chiral limit) 
is less than 0.2fm$^{-3}$ there, which is much less than the baryon density 
we are interested in ($\rho\geq 4\rho_0$). 
Besides that, there is another reason why $\pi-N$ and
$\pi-K$ interaction are not so important in our discussion. 
Experimental data shows that the $s$-wave interaction hardly contributes, 
at low momentum, in isospin-symmetric matter, which can 
be also explained by the chiral-symmetry argument. 
Hence the leading contribution comes from the $p$-wave interaction. 
Its contribution to the effective energies of baryons or kaons in the medium, 
which is essentially proportional to the square of pion momentum or that of 
kaon momentum, then enters through the
modification of the dispersion relations (Eqs. (7),(9)). Its magnitude for the
effective masses of kaons and baryons is proportional 
to $T^4$ in the chiral limit, and then much suppressed for $T<
m_\pi$; e.g. $\delta M_N < $ several tens MeV by way of the
scattering-volume approximation, which should be compared with the one 
from RMFT, $\delta M_N \sim$ several hundreds MeV even at $\rho_0$. 
>From these discussions we 
may safely say that thermal effects play little role in our discussion and
the effects of RMFT are dominant at temperature below $m_\pi$, that is, 
the density effects overwhelm the temperature effects in short.

In summary we have discussed the possibility of \Kp(\Ko)-condensation
in high-density baryon matter
and suggested
that kaon condensation is caused by the difference of the vector
mean-fields for
nucleon, lambda and kaon; this difference becomes linearly larger with  
density.
Thus the possibility of this condensation depends on the strength of
their vector mean-fields. 
We have found a simple criterion for kaon condensation within RMFT
(Eq.(13)). If parameters satisfy this condition, we can expect kaon
condensation at some density. The value of the critical density depends
on the value of the
nucleon effective mass  $M^*_N$ at the saturation density. We have
found $4-16\rho_0$ for the critical density within our parameter sets, 
PM1 and PM4.
The typical value of effective mass is empirically known as 
$M^*_N/M_N = 0.55 - 0.7$
\cite{Rufa,boguta,Hama,DBHF,Qing},
and our parameter-sets are consistent with them.
Furthermore
the small LS-splitting of lambda hypernuclei and the very tiny Hartree 
contribution to the nucleon vector mean-field suggests the 
small vector mean-field for lambda.

There remains a problem of the possibility of  \Kp(\Ko)-condensation
in high-energy heavy-ion collisions. There is controversy at present 
about how a
system reaches thermal equilibrium and to what extent density and/or 
temperature are raised. 
Recent numerical simulations have shown that baryon density 
$\rho/\rho_0 = 7 - 10 $ can be achieved  by the high-energy heavy-ion 
collisions with several tens GeV/u like 
in the AGS energy region \cite{AGS1,AGS2}.
In this case the system is expected to be quasi-equilibrilium for the
duration of 4 - 8 fm/c
with the temperature $T \approx 120$MeV \cite{AGS1}, which is still below 
the pion mass. Hence \Kp(\Ko)-condensation is very plausible to be generated 
in the high-density regime.
In the heavy-ion collisions around hundreds GeV/u like in the SPS
energy 
region it depends on the model whether the system is stopped
and equibrilium is realized\cite{gyulassy}; e.g.,
temperature becomes $T \approx 140$MeV in the RQMD simulation 
\cite{RQMD-T} which may be a marginal temperature for our discussion 
to be applied. 
\Kp(\Ko)-condensation
, once occurs in course of relativistic heavy-ion collisions
around the AGS energy region or higher, would 
give rise to a new phenomenon for dilepton production \cite{shin}.
Of course we must wait for numerical simulations in order to conclude
whether this phenomenon can be observed in the actual heavy-ion collisions.
One promising way for this purpose is 
the RBUU approach \cite{LiKo,RBUU,TOMO2}, 
but we need to introduce the momentum-dependence for the mean-fields in this 
high-energy region \cite{TOMO2}.
It may be a very hard task in the light of the present computer power.

In this letter we have taken into account only the essential degrees
of freedom for kaon condensation in symmetric nuclear matter.
They  would be sufficient to discuss the onset of the condensation \cite{shi},
while we must consider other strange particles besides $\Lambda$ and
kaon to get a realistic description of the condensed phase. 
Full argument including $SU(3)$ octet baryons and mesons will be given in a 
separate paper \cite{shi}.

Two of authors (T.T and H.F.) are grateful to the Institute for
Theoretical Physics at the University of Adelaide for its hospitality
during the Workshop on Quarks, Hadrons and Nuclei. 
One of us (T.T.) wishes to thank Prof. Thomas and Dr. Tsushima for useful 
discussions. 
One of us (T.M.) wishes to thank Prof. Motoba and Dr. Nara for useful 
information on hypernuclei and high-energy heavy-ion collisions. 
This work
was supported in part by the Japanese Grant-in-Aid for Scientific
Research Fund of the Ministry of Education, Science and Culture 
(06640388, 08640369).

\newpage

{\Large Figure Captions}

\bigskip

\begin{itemize}

\item[Fig. 1]
Density-dependence of ${\Delta \mu}$.
The dotted, solid, chain-dotted, dashed and thick dotted lines indicate
the results in PM1-L1, PM1-L2, PM4-L1, PM4-L2 and PM1-free, respectively.

\item[Fig. 2]
Density-dependence of the total energy per nucleon (a),
the ratio of the effective masses to the bare masses for nucleon and kaon (b), 
and the kaon fraction (c).
The dashed and thick dotted lines indicate the results for PM1-L2 and PM4-L2,
respectively.
In the second column (b) the upper two lines show the effective mass of kaon
and the lower ones show that of nucleon.
The thick and thin lines in (a) denote EOS under kaon condensation
and normal phases, respectively. 

\item[Fig. 3]
Parameter-dependence of the critical density.
The $x$-axis corresponds to the ratio $x_\omega = \gvL/\gv$.
The solid and dashed lines indicate the results for PM1 and PM4,
respectively.

\end{itemize}

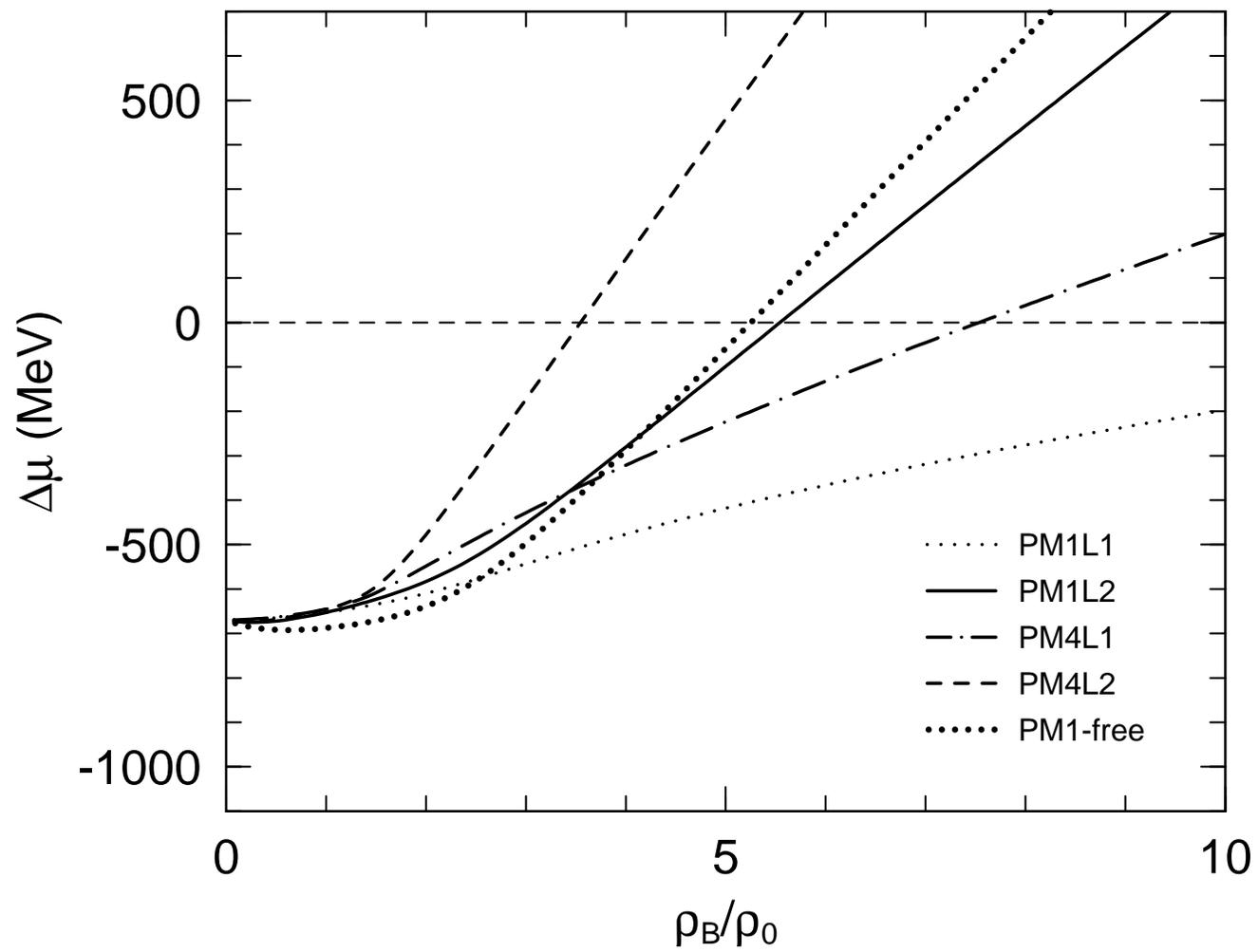
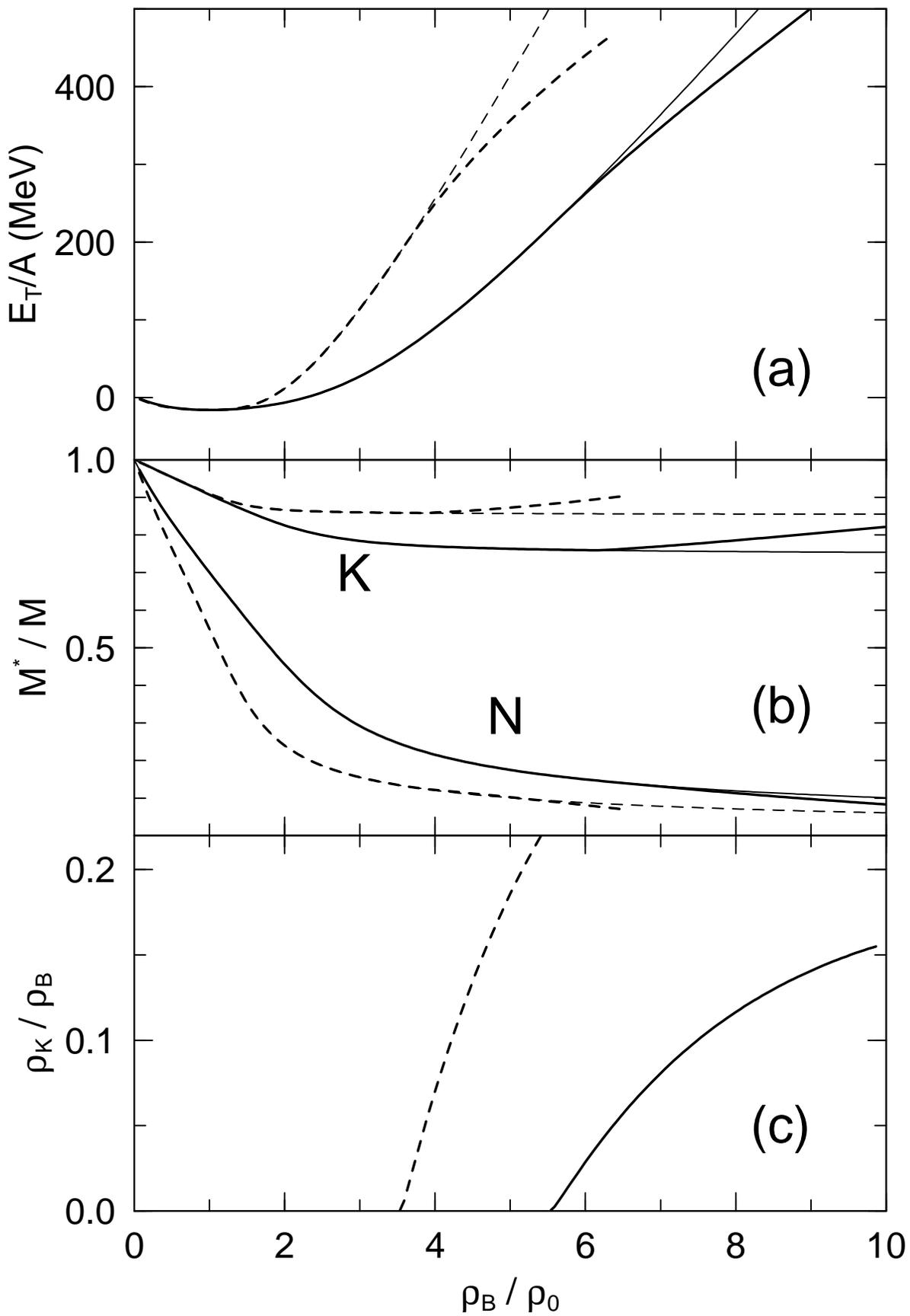
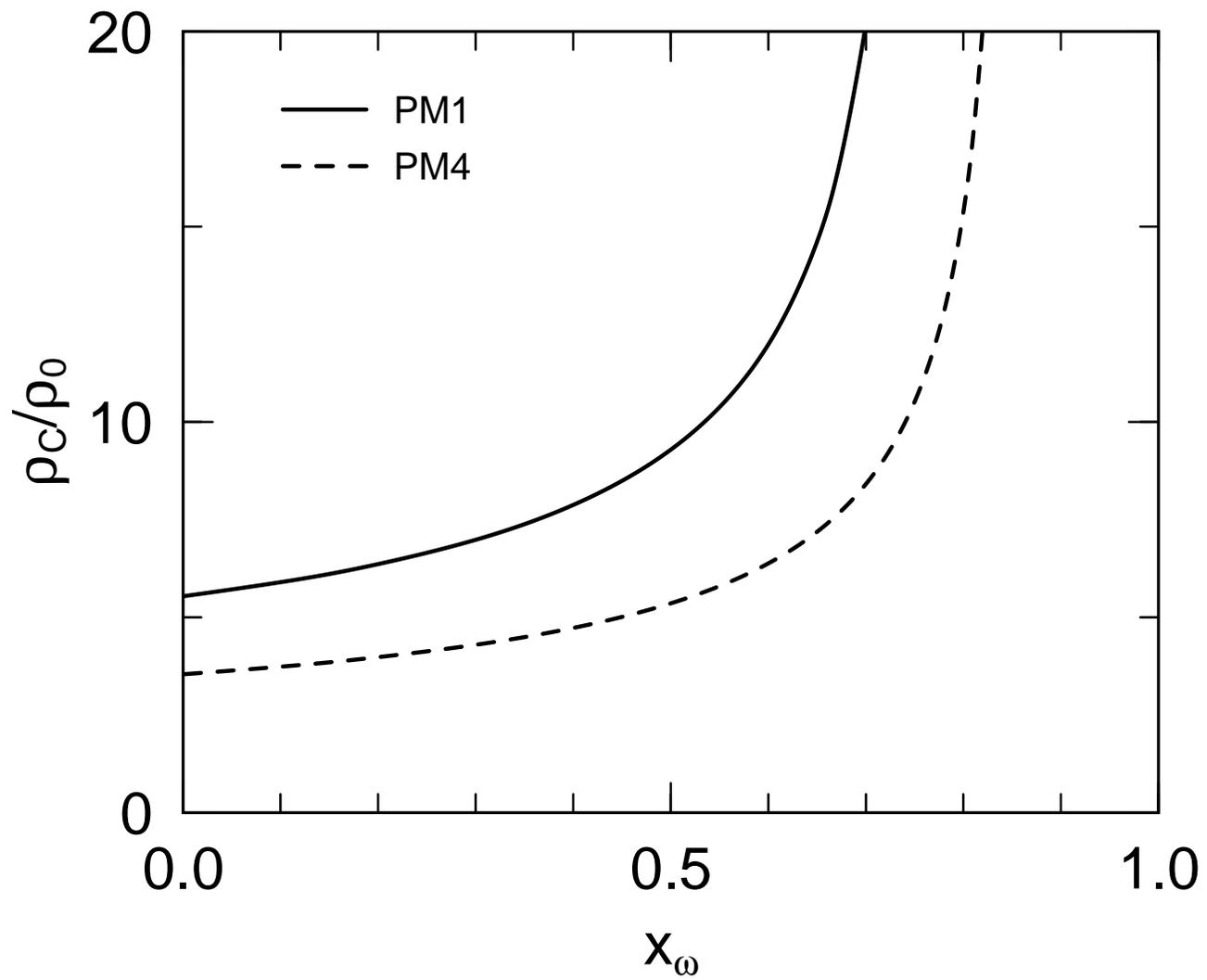
\end{document}